\documentclass[prb, twocolumn, superscriptaddress, showpacs,floatfix]{revtex4}  
\usepackage{hyperref}
\usepackage{amssymb}
\usepackage{amsmath}
\usepackage{float}
\usepackage{bbm}
\usepackage{graphicx}
\usepackage{epsfig}
\usepackage{epstopdf}
\usepackage[usenames]{color}

\begin{document}

\title{Covalency, double-counting  and the metal-insulator phase diagram in  transition metal oxides}

\author{Xin Wang}
\affiliation{Condensed Matter Theory Center, Department of Physics, University of Maryland, College Park, MD 20742, USA}

\author{M. J. Han}
\altaffiliation[Current address:]{ KAIST, South Korea
}
\affiliation{Department of Physics, Columbia University, 538 West 120th Street, New York, NY 10027, USA}

\author{Luca de{'} Medici}
\affiliation{Laboratoire de Physique des Solides, Universit\'e Paris-Sud, CNRS, UMR 8502, F-91405 Orsay Cedex, France}

\author{Hyowon Park}
\affiliation{Department of Applied Physics and Applied Mathematics and Department of Physics, Columbia University, New York, New York 10027, USA }

\author{C. A.  Marianetti}
\affiliation{Department of Applied Physics and Applied Mathematics, Columbia University, New York, New York 10027, USA }

\author{Andrew J. Millis}
\affiliation{Department of Physics, Columbia University, 538 West 120th Street, New York, NY 10027, USA}

\date{\today}


\begin{abstract}
{Dynamical mean field theory calculations are used to  show that for late transition-metal-oxides  a critical variable for the Mott/charge-transfer transition is the number of $d$-electrons, which  is determined by charge transfer from oxygen ions. Insulating behavior is found only for a narrow range of d-occupancy, irrespective of the size of the intra-d Coulomb repulsion.  The result is useful in interpreting 'density functional +U' and 'density functional plus dynamical mean field' methods in which additional correlations are applied to a specific set of  orbitals and an important role is played by the 'double counting correction' which dictates the occupancy of these correlated orbitals.    General considerations are presented and are illustrated by calculations for two representative  transition metal oxide systems: layered perovskite $Cu$-based ``high-$T_c$ materials, an orbitally non-degenerate electronically quasi-two dimensional systems, and pseudocubic rare earch nickelates, an orbitally degenerate electronically three dimensional system.  Density functional calculations yield d-occupancies very far from the Mott metal-insulator phase boundary in the nickelate materials, but  closer to it in the cuprates, indicating the sensitivity of theoretical models of the cuprates to the choice of double counting correction and corroborating  the critical role of lattice distortions in attaining the experimentally observed insulating phase in the nickelates.  }
\end{abstract}

\maketitle

\section{Introduction}
The strong electronic correlations characteristic of transition metal oxides pose a problem of long-standing interest \cite{Mott49,Imada98} and great current importance \cite{Held06,Kotliar06,Mannhart08,Muller08,Ament10}.  In transition metal oxides the important low energy electronic states are derived from the transition metal d-orbitals. In a simple ionic picture the ground state corresponds to a $d^n$ configuration for each transition metal ion, with d-electron number $n$ depending on material.  The correlation energy cost involved in changing the d-occupancy is conventionally denoted by $U$,  is defined as  $U=E[d^{n+1}]+E[d^{n-1}]-2E[d^n]$ and is typically large. In a ground-breaking paper Mott \cite{Mott49} argued that  if  $U$ were large enough relative to the bandwidth then electron repulsion  would prohibit conduction, at least at integer valence.

Mott's concept was refined in an important way by Zaanen, Sawatzky and Allen \cite{Zaanen85} who noted that correlated electron materials of interest typically  contain ligands (for example the $O$ atoms in transition metal oxides) which are  close enough in energy  to the chemical potential that one must also consider the transfer of an electron from a ligand to a transition metal site.  Ref. ~\onlinecite{Zaanen85}  defined the  charge transfer energy $E_{CT}=E[d^{n+1}]-E[d^{n}]-\varepsilon_p$ with $-\varepsilon_p$ the energy needed to create a hole in a ligand orbital  and demonstrated  that if $U$ were large but $E_{CT}<U$ then the physics would be controlled by $E_{CT}$  with $U$ playing a relatively minor role.  Implicit in this argument is a neglect of correlations on the oxygen sites; the justification is that the number of ligands is large enough and the hole density on the ligands small enough that configurations involving doubly occupied ligand sites may be neglected. The resulting 'charge transfer insulator' paradigm is  central to the conventional understanding \cite{Imada98} of the physics of  transition metal oxides.

While density functional band theory (DFT) is the work-horse of materials science  \cite{Jones89}, it does not capture the physics of the Mott/charge-transfer insulator transition: theoretical study of charge transfer insulators and other correlated electron materials requires methods which include additional correlations. Such ``beyond DFT'' methodologies require the identification of a set of orbitals whose correlations are to be treated more accurately, an ansatz for the extra interactions operating among these orbitals, a method of solution of the correlation problem and a prescription for embedding the correlated orbitals into the full electronic structure. In the beyond DFT studies of  transition metal oxide materials published to date, the correlated orbitals are taken to be all or a subset of the transition metal d-states, the extra interactions are matrix elements of the Coulomb interaction, projected onto the relevant part of the d-manifold and screened by the other degrees of freedom in the solid, and the method of solution is either a Hartree approximation, which gives the widely used ``DFT+U'' \cite{Anisimov91} approach or dynamical mean field theory (DMFT), which gives  the  ``DFT+DMFT'' \cite{Held06,Kotliar06} approach. The crucial feature of the embedding is the 'double counting correction' \cite{Anisimov91,Amadon08,Karolak10}, which removes  correlation contributions from the DFT single-particle energies. The double-counting correction plays a particularly significant role in charge transfer insulators because it affects the value of the p-d energy difference which is crucial to the  physics. However, the double counting correction is not well understood theoretically: several different choices are commonly used\cite{Anisimov91,Amadon08,Karolak10} but no exact prescription is known.

In this paper we argue that the double counting problem should be recast as a problem of determining the occupancy, $N_d$ of the correlated orbitals. Expressing the problem in this way reveals a dramatically simplified picture of the metal-insulator transition in which the only important  variables are the correlation strength and $N_d$. The complicated real-materials effects of ligand dispersion and ligand-transition metal hybridization and the various theoretical issues related to  double counting and other embedding aspects are seen to affect the phase boundary and spectra only via their effects on $N_d$, as long as well localized orbitals are chosen.  Expressing the problem in this way also highlights an important but perhaps under-appreciated aspect of the DFT+DMFT procedure  \cite{Held06,Kotliar06}: the physics predicted by this method is to a large degree controlled by the choice made for the double-counting correction.  The rest of this paper is organized as follows. Section~\ref{Methods} outlines our theoretical approach, section~\ref{Results} presents results obtained using beyond DFT methods for rare earth nickelates and cuprates, and section~\ref{Discussion}  gives a discussion and conclusion. 

\section{Theoretical Approach \label{Methods}}
\subsection{Formalism}
In this paper we use the DFT+DMFT methodology \cite{Georges04,Held06,Kotliar06,Amadon08}. In this procedure we define in each unit cell $j$ (central position $R_j$)  a set of ``correlated orbitals'' $\phi_j^a(r-R_j)$ (we discuss the mechanics of defining the $\phi^a$ below).  Then, following the formal procedure outlined in Ref.~\onlinecite{Kotliar06} one   constructs a functional $\Gamma[\rho,\hat{G}_{loc}]$ of the charge density $\rho(r)$ and an auxiliary local  Green function $\hat{G}_{loc}$. $\Gamma$ may be formally defined in terms of the standard Kohn-Sham density functional and the Luttinger-Ward functional of $\hat{G}_{loc}$. Extremizing the functional leads  to the Kohn-Sham-Dyson equation relating the self energy $\hat{\Sigma}$ and Green function $\hat{G}$, which are matrices in the full single-particle Hilbert space:

\begin{align}\label{}
\hat{G}(\omega)=\left(\omega\mathbf{1}-\hat{H}_{ks}-\hat{\Sigma}(\omega)\right)^{-1}
\label{ksd}
\end{align}
The self energy is obtained as the functional derivative of the Luttinger-Ward functional with respect to $\hat{G}_{loc}$ and the equations are closed by  a self-consistency equation relating $\hat{G}_{loc}$ to the components $<\phi^a|\hat{G}|\phi^b>$ of $\hat{G}$. For details see for example Ref.~\onlinecite{Kotliar06}.  

We  construct and solve the Kohn-Sham Hamiltonian using the generalized gradient approximation (GGA) (typically via  the Vienna Ab-initio Simulation Package (VASP) \cite{Kresse1993558,Kresse199414251,Kresse199615,Kresse199611169} but in selected cases using the WIEN2K package \cite{Wein2K}). 

To obtain the self energy we use the single-site dynamical mean approximation \cite{Georges96}. This is generally accepted as containing the essence of the Mott and charge-transfer metal-insulator transition physics and is widely used in DFT+DMFT studies.  Multi-site (``cluster'') methods are too computationally expensive to be practical for DFT+DMFT calculations (particularly for orbitally degenerate situations where Hunds interactions are important). In model systems, cluster methods have been shown to  provide a more refined picture  but confirm that  the single-site approximation captures the important high energy physics of the metal-insulator transition  \cite{Gull10}. In particular a reasonable working definition of whether or not a material is ``strongly correlated'' is whether the stoichiometric compound is on the metallic or insulating side of the single-site DMFT phase boundary \cite{Comanac08,Weber10}. and is widely used in DFT+DMFT studies of real materials.

In the single-site dynamical mean field approximation we write

\begin{equation}
\hat{\Sigma}\rightarrow\hat{\Sigma}^{DMFT}-\hat{E}^{dc}
\label{DMFTapprox}
\end{equation}
with
\begin{equation}
\hat{\Sigma}^{DMFT}=\sum_{jab}\left|\phi^a_j\right>\Sigma^{ab}(\omega)\left<\phi^b_j\right|
\label{SigmaDFMT}
\end{equation}
and $\hat{E}^{DC}$ the frequency-independent double counting correction which plays a crucial role in our subsequent considerations.  The need to include this term may be seen from the argument that the underlying band theory includes some aspects of the many-body physics within a static approximation; the additional interactions that go into the computation of $\hat{\Sigma}$ in effect count these terms twice so a correction is needed. In practical terms, the importance of the double-counting term is that any many-body computation of $\hat{\Sigma}$ will lead to a static Hartree-Fock contribution to $\hat{\Sigma}$ which will substantially shift the single particle levels. The double counting correction acts to compensate for this shift.  How to specify the double counting term $\hat{E}^{DC}$ is an important open problem in materials theory. Different prescriptions   have been proposed \cite{Anisimov91,Amadon08,Karolak10}, but no clear consensus has emerged. We therefore explore a range of double countings.

The self energy matrix components $\Sigma^{ab}$ are obtained from the solution of a quantum impurity model which  is specified by the additional beyond band theory  interactions discussed in more detail below and by a  hybridization function $\Delta^{ab}$ obtained the self-consistency condition,which in the single-site approximation  relates $G^{ab}(j,j,\omega)$, the a-b components of the unit-cell local  $G$, to the same components of the  Green function $G_{loc}$ of the quantum impurity model. Explicitly (the inversion is in the orbital (a-b) space)
\begin{equation}
\Delta^{ab}(\omega)(j,j,\omega)=\omega\mathbf{1}-\Sigma^{ab}(\omega)-\left[G^{ab}(j,j\omega)\right]^{-1}
\label{SCE}
\end{equation}
For orientation we remark that the matrix of d-level energies determined e.g. from the Wannier fit to the band structure is equivalently given as the infinite frequency limit of $\Delta$.

We solved the quantum impurity model using the hybridization expansion method \cite{Werner06,Werner06b,Gull11}; some results were cross-checked via the exact diagonalization solver \cite{Caffarel94,Capone07}.

Finally we note that to complete the formal solution one must compute the charge density from $\rho(r)=\int d\omega/\pi f(\omega)Im G(r,r,\omega)$ and ensure that this charge density is used to obtain the Kohn-Sham Hamiltonian. We discuss issues of 'full charge self-consistency' below.

\subsection{Choice of correlated orbitals and interaction}

In this paper we focus on transition metal oxides. Following common practice in the literature, we assume that the important beyond band theory correlations involve atomic-like transition metal d-orbitals. These are not uniquely defined. The ambiguity has a physical origin:  in a  solid the overlap of tails of wavefunctions defined around different atoms, along with the need to introduce additional states to describe the interstitial regions means that  there is no unique definition of an atomic orbital. One must seek a practical definition which corresponds reasonably closely to the physical/chemical intuition of an atomic orbital.    In most of our calculations we construct the d-orbitals using the maximally localized Wannier function procedure of \cite{Marzari97,Souza01,Mostofi08}  with a wide energy window (typically $-10$ to $3$ eV with the zero of energy chosen as the Fermi level) which spans the entire p-d manifold of states. Use of the wide energy window is essential to obtain reasonably atomic-like d-orbitals.  In selected cases we used the projector method~\cite{Amadon08,Haule:10} with a correspondingly wide energy window. 

For the interactions among the d-orbitals we assumed
\begin{equation}
H_d^{int}=\frac{U}{2}{\hat N}^i_d\left({\hat N}^i_d-1\right)+H_J
\label{Hddef}
\end{equation}
Here, ${\hat N}^i_d=\sum_{a\sigma}d^\dagger_{ia\sigma}d_{ia\sigma}$ is  the number operator  for $d$ electrons on site $i$ and $H_J$ representing the additional `Hunds-rule' interactions which give the multiplet structure at fixed $N_d$ (for explicit forms see e.g. Ref.~\onlinecite{Imada98}).  The  ``charging energy'' $U$ couples to the total on-site charge, which is strongly electrostatically coupled to the surrounding ions so    is renormalized significantly by screening. The appropriate values for different materials are not well determined theoretically  although interesting results have appeared \cite{Aryasetiawan06}.  We therefore consider a range of $U$ values spanning the range estimated from experiment.  The  configurations coupled by  $H_J$ interact with the rest of the solid only via the electric quadrupole field for which screening is negligble, so these terms are well approximated by their gas-phase values, in agreement with calculation \cite{Aryasetiawan06}.

\subsection{Relation to Phenomenological Models}

In the literature \cite{Zaanen85,Imada98} the late transition metal oxides are often described by a phenomenological 'p-d' model, tight-binding formulation in which the Hamiltonian includes correlated (``d'') orbitals with appropriate on-site interactions,  ligand (oxygen) orbitals (assumed uncorrelated because the hole density is small)  and the relevant hybridizations. It may be written as
\begin{equation}
H_{pd}=\sum_{ia\sigma}\varepsilon_{d,ia\sigma}d^\dagger_{ia\sigma}d_{ia\sigma}+H_{int}+H_{hyb}+H_{ligand}
\label{Hpd}
\end{equation}
The maximally localized Wannier representation of the band structure in the energy window $-10$ to $3eV$ can be thought of as a derivation, from a first-principles band structure, of the hybridization parameters $H_{ligand}$ and $H_{hyb}$ describing the embedding of the correlated orbitals into the broader electronic structure. The actual band structure gives a relatively involved form for these terms (see, e.g. \cite{Weber10} for cuprates) whereas many literature papers have considered simpler models where for example oxygen-oxygen hopping is neglected or treated in a simplified manner. However, the important qualitative and quantitative aspects of the many-body physics have been shown to be insensitive to the details of $H_{ligand}$ and $H_{hyb}$ \cite{Wang11}, at least within the single-site DMFT approximation. The reason, in essence, is that the band structure enters the many-body physics only via the hybridization function $\Delta$ (Eq.~\ref{SCE}), in which most of the details are averaged over. 

As shown first by Zaanen et. al, the  crucial parameter is the difference $E_{CT}=\varepsilon_d-\varepsilon_p$ between an average ligand on-site energy $\varepsilon_p$ and  the orbitally averaged $d$ level energy $\varepsilon_d=\sum_a\varepsilon_d^a/n_{orb}$. If $E_{CT}$ is small or negative  the d-level lies inside the p-band and the hybridization ensures that model is metallic even at large $U$, whereas if $E_{CT}$ is large and positive the d-bands are well separated from the p-bands and an insulating state may occur if $U$ is large enough.  

While the qualitative dependence of the physics on $E_{CT}$ is clear, four ambiguities arise in practice. First, if oxygen-oxygen hopping is important (as is the case in known transition metal oxides) the p-states are spread over a wide energy range even before hybridization to the d is considered, but only some portion of these states are hybridized to the d-levels. Thus the `average ligand on-site energy' referred to above is not clearly defined in realistic cases.  Second, the double counting correction enters $H_{pd}$ as a shift of the d-level and therefore enters $E_{CT}$ directly. Thus the value  of the double counting correction directly affects the physics and indeed  uncertainties in this parameter are a significant source of uncertainty in the theoretical results. Third, the many-body interactions $H_{int}$ will shift the physical d-level (as can be easily seen on the Hartree level); thus the physical value of $E_{CT}$ also depends on $U$. Finally, the self-consistent nature of the DFT+DMFT procedure means that the band parameters (in particular the $\varepsilon_d$) are themselves affected by the solution to the many-body problem. 

One important goal of the present paper is to show that these ambiguities to a large extent disappear if the theory is parametrized not by $E_{CT}$ but by the d-occupany $N_d$.

\subsection{d-occupancy $N_d$}

In the rest of this paper an important role is played by the d-occupancy $N_d$. This is a theoretically constructed quantity, defined in terms of a representation $\phi_d^a(r)$ (here we define the origin of coordinates to be centered on a transition metal ion and consider only one unit cell) obtained by one of the methods described above. The definition is
\begin{eqnarray}
N_d&=&\sum_{a,\sigma}\int\frac{d\omega}{\pi}f(\omega)\int d^3rd^3r^{'}
\\
&&~~\times Im\left[\left( \phi_d^a(r)\right)^* G_\sigma(r,r^{'},\omega)\phi_d^a(r^{'})\right]
\nonumber
\label{Nddef}
\end{eqnarray}
Note that in using Eq.~\ref{Nddef} it is important to work with properly normalized $d$ -orbitals $\phi_d$. The maximially localized Wannier method automatically provides these, but  when using projector methods one must typically normalize the resulting d-orbitals. 

In the tight-binding p-d model of Eq.~\ref{Hpd} Eq.~\ref{Nddef} becomes 
\begin{equation}
N_d=\sum_{a\sigma}\langle d^\dagger_{ia\sigma}d_{ia\sigma}\rangle
\label{Ndpd}
\end{equation}
where   the wave functions are implicitly included via their effect on the hybridization parameters.

We shall see in the analysis to follow that for interactions $U$ in the physically reasonable range for late transition metal oxides, and for well localized $\phi_d$, all of the details of ligand band structure, of choice of double counting, of full charge self consistency and of DFT+DMFT vs p-d model  and  of the precise definition (Wannier vs projector) of the $\phi_d$ important only insofar as they affect $N_d$. Models with the same $N_d$ give, to a very good approximation,  the same metal-insulator physics. 

At this point one remark is important. The d-orbital is $5$-fold degenerate. In many cases of physical interest only a few of the orbitals are relevant (for example, the $d_{x^2-y^2}$ orbital in high-$T_c$ cuprates); ligand field effects mean that  others are fully filled or empty.  The $d$-occupancy which is important is that of the relevant partially filled orbitals. The other ``filled'' orbitals are of course hybridized to the ligands and in any band theory will have occupancies which deviate slightly from integer values. The precise occupancies of these orbitals are not important, but the presence or absence of these orbitals in the definition of $N_d$ may shift phase boundaries. 

As noted above, $N_d$ is a theoretical quantity whose definition involves a choice of an orbital.   The quasi-universal dependence of physics on $N_d$ which we will demonstrate below however suggests that it is interesting to attempt to relate the calculated  $N_d$ to experiment and to other calculations. Such a relation is of necessity not precise, but may be informative.  

Density functional band theory (without any additional correlations) provides an estimate of $N_d$ via one of the constructions outlined above.  In principle, if the exact density functional were known, DFT would produce the exact charge density $\rho(r)=\int\frac{d\omega}{\pi}f(\omega)\int d^3r Im G_\sigma(r,r,\omega)$, and existing approximate implementations are believed \cite{Jones89} to provide very accurate representations of $\rho(r)$. This does not guarantee that the DFT estimate of $N_d$ (which as seen from Eq.~\ref{Nddef} depends on $G$ at two different spatial points) is exact even given the exact density functional, but the small spatial extent of the d-orbitals and the general success of DFT describing basic charge properties suggests that it is reasonable to expect that the physical $N_d$ are not too far from the DFT estimates. 

Experimental estimation of the $d$ valence is  possible via analysis of  transferred hyperfine couplings \cite{Walstedt01} or   magnetic form factors  \cite{Walters09}. These experiments have indicated very strong covalency effects in cuprates. Resonant X-ray scattering \cite{Peets09,Ament10} can detect the density of holes on oxygen. Further, spectroscopic measurements can in some cases identify the $\varepsilon_{p,d}$ energies, permitting more quantitative comparison to theory. These issues will be discussed further below.

\section{Results \label{Results}}

\subsection{Formulation}

Our specific calculations are performed for $La_2CuO_4$ and $LaNiO_3$.  We consider only the idealized crystal structures; simple cubic $ABO_3$ perovskite  for $LaNiO_3$ (lattice constant $3.8366 \AA$ ; simple tetragonal (T) structure for $La_2CuO_4$ (lattice constants  $3.83421\AA$ in plane and $13.1831\AA$ c-direction with the z position of apical oxygens  $0.186/0.814$ \% of the c-axis lattice parameter.  (The lattice constants and the atomic positions were determined  using GGA relaxation calculations and differ slightly from those used by other groups; the differences are not important for present purposes).  For most of our calculations on $La_2CuO_4$ we treat one correlated $d$ orbital per Cu site ($d_{x^2-y^2}$) with DMFT; the other $d$ orbitals are treated approximately, using  Hartree-Fock or density functional theory.  For a few parameter values we checked the one-dynamically correlated orbital approximation  by solving the full five-orbital problem with DMFT (retaining only the density-density terms in the interaction).  We found no significant difference in the results of the metal-insulator transition.  For $LaNiO_3$ we chose two correlated orbitals per Ni site, representing the $x^2-y^2$ and $3z^2-r^2$ states.  Recently Deng and collaborators used a full five orbital model (also retaining only the density-density terms)  to study the nickelates \cite{Deng12}; their results (apart from some details of excitation from the $t_{2g}$ bands)  are again essentially the same as those obtained from the two-correlated orbital model.

To define the correlated orbitals and determine the hybridization function  we typically used GGA calculations with VASP to define the band structure and  maximally localized Wannier functions defined over an energy range of -9$\sim$3eV  for $La_2CuO_4$ and -8$\sim$4eV for $LaNiO_3$ to fix the bare hybridization function and  the choice of $d$-orbital.  We also cross-checked the $La_2CuO_4$ results with the Wien2K/Projector method~\cite{Wein2K,Haule:10} using the same energy window of -9$\sim$3eV for the consistency.  In our calculations for both materials we followed the literature by   using an orbital-independent double-counting correction, so $\hat{E}^{DC}$ is the unit matrix in the space of dynamically correlated $d$ orbitals.  We considered a range of $E^{DC}$, which we parametrized by the resulting $N_d$.  In most of  our calculations we kept the Wannier fits (i.e. the bare hybridization function)  constant as we varied $E^{DC}$.  To test this approximation for $La_2CuO_4$ we also performed fully charge self-consistent DFT+DMFT calculations using the Wien2K/projector scheme.

For beyond band theory interactions we choose the Slater-Kanamori/Hubbard type Hamiltonian \cite{Imada98} parametrized by the on-site Coulomb interaction  and exchange term. In the literature and the different beyond theory methods different definitions are used. In the VASP/Wannier+DFT calculations  Coulomb interaction $u$ and the Hund's coupling $j$  are defined in terms of
Slater parameters: $F^0$, $F^2$, $F^4$ as
\begin{eqnarray}
u &=& F^0+(4/49)\cdot (F^2+F^4)\\
j &=& (5/98)\cdot (F^2+F^4)
\end{eqnarray} 
The Wien2k/Projector scheme always treats an entire five d orbital interation and the full interaction tensor is also given in terms of the Slater integrals, $F^0$, $F^2$, and $F^4$. For sufficiently symmetric situations such as those considered in this paper the interaction can be parametrized in terms of interactions $U$ and $J$ defined as
\begin{eqnarray}
U &=& F^0 \\
J &=& (1/14)\cdot (F^2+F^4)
\end{eqnarray}
Therefore, $u$ and $j$ defined in VASP/Wannier are related to $U$ and $J$ in Wien2k/Projector:  $u=U+(8/7)J$, $j=(5/7)J$.  In this paper, we show the results of only $U$ and $J$ parameters.  In most of our calculations we used $J=0.7eV$ for both materials and varied $U$  to map out  the metal-insulator phase diagram.  For $La_2CuO_4$, we also present some results for $J=0$.

We solved the impurity model using the fully rotationally invariant hybridization expansion \cite{Werner06b,Gull11}  except that for the five-orbital $La_2CuO_4$ calculations we used the 'segment method' \cite{Werner06,Werner06b,Gull11} which means we omit the exchange and pair-hopping interactions.  This approximation is commonly used in the literature (see, e.g. \cite{Deng12}) Our cuprate calculations were performed at temperature $T=0.02eV$ and our nickelate calculations mainly at $T=0.1eV$ but in selected cases cross-checked at temperatures $T=0.05eV$. While these temperatures are high compared to room temperature they are very low compared to the energy scales relevant here.  For the one-orbital $La_2CuO_4$ calculations we cross checked some of the results  using the exact diagonalization method \cite{Caffarel94,Capone07}

Previous work on $La_2CuO_4$  \cite{Weber10,Wang11}  had used a published \cite{Weber10} tight binding parametrization of a  band structure  which led to an $N_d\approx 1.65$. Our calculations, in agreement with unpublished work of Haule et al, lead to a   band theory estimate $N_d\approx 1.45$ rather than the $\approx 1.63$ presented previously.  The difference in band structure leads to small differences from previous work. 

\subsection{Metal-Insulator Phase Diagram}

\begin{figure}[b]
\begin{center}
\includegraphics[angle=0.0, width=0.9\columnwidth]{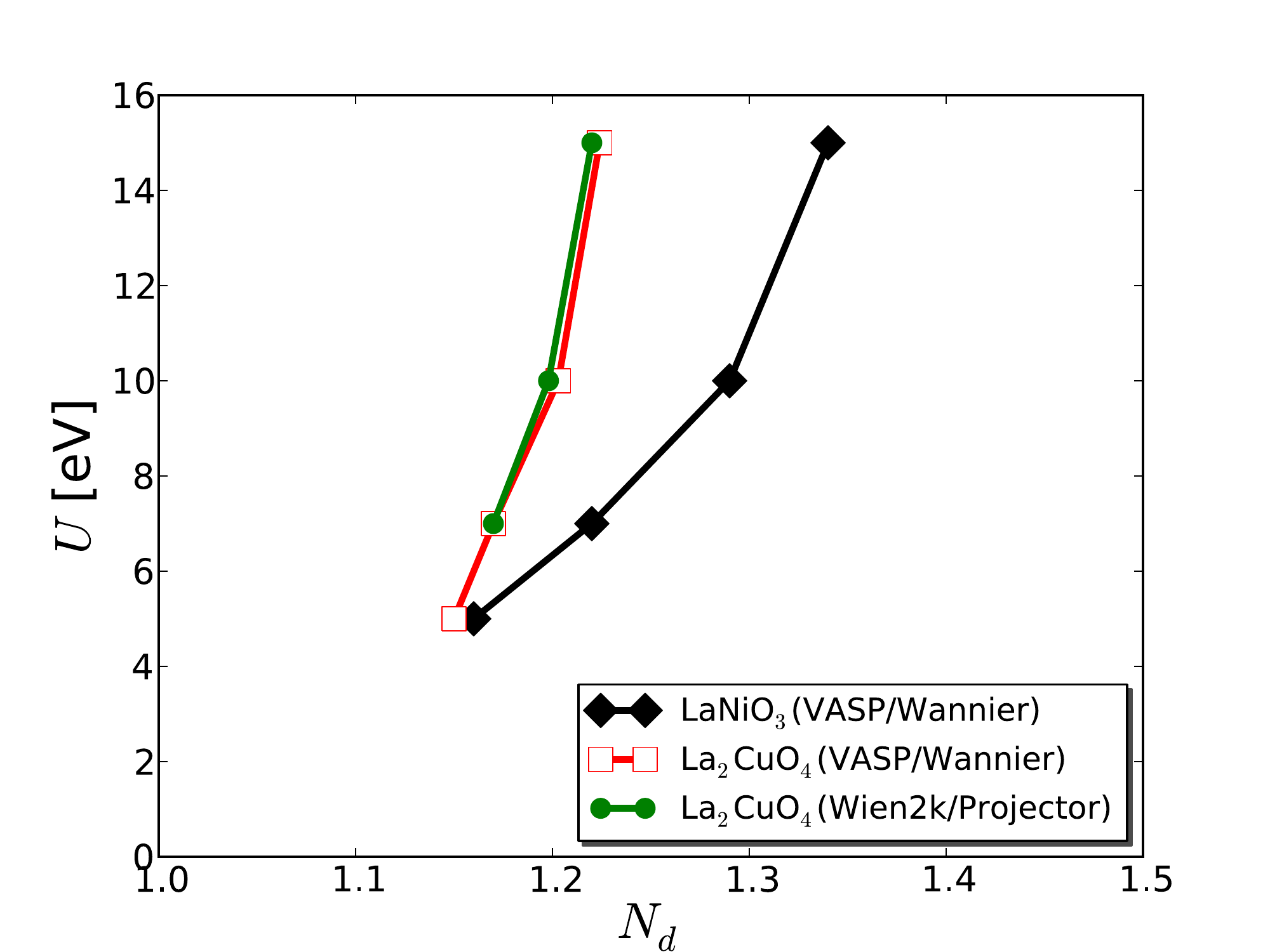}
\caption{(Color online) Metal-insulator phase diagram in plane of charging energy $U$ and d-occupancy $N_d$ (measured relative to the $d^8$ (cuprates) or $d^6$ (nickelates) core) indicating limit of stability  of insulating phase  $N_{d,c1}$ for $La_2CuO_4$ (calculated from the one-correlated orbital model with VASP/Wannier (red empty square) and five-correlated model with Wien2K/projector (green filled circle) methods) and $LaNiO_3$ (calculated from the two correlated orbital model with VASP/Wannier (black filled diamond)).  $J$=0.7eV is used for all calculations. The $U=0$ $N_d$ values of $La_2CuO_4$ are 1.45 (Wien2K/projector) and 1.47 (VASP/Wannier). The $U=0$ $N_d$ value of $LaNiO_3$ is 2.18 (VASP/Wannier).
}
\label{pdnd}
\end{center}
\end{figure}

Fig.~\ref{pdnd} plots the metal insulator phase diagrams calculated for the two materials, in the plane of charging interaction $U$ and d occupancy $N_d$. Different values of $N_d$ are obtained by varying the double counting energy $E^{DC}$ keeping the rest of the band structure fixed. When plotted in terms of $E_{CT}$ the results for $La_2CuO_4$ are consistent with results  previously presented  in the literature for cuprates \cite{Zolfl98,Wang11}.   

In single-site dynamical mean field theory the metal insulator transition is first order and is characterized by two spinodal lines: one at which  the insulating phase loses stability and one at which the metallic phase loses stability \cite{Georges96}.  Because  the limit of stability of the metallic phase has a strong temperature dependence which  is computationally very expensive to capture, we present here the limit of stability of the insulating phase $U_{c1}/N_{d,c1}$ .

At small $U$ one might expect that the phase boundaries of extrapolate linearly towards $U=0$, $N_d=1$ because as $N_d \rightarrow 1$ the $d$ bands become far removed from the $p$ bands so both the $d$ bandwidth and $N_d-1$ $\sim 1/\Delta$ (there will be some small d-d hopping which will make U extrapolate to some small non-zero value). We have not pushed our calculations into this regime because it is unphysical for the transition metal oxides. Accessing the small-U phase boundary by tuning parameters so that $N_d$ is moved  closer to $1$  corresponds to placing the d-bands at such a high energy relative to the p-bands  that they would be mixed with transition metal $t_{2g}$ bands, rare earth $f$ and $d$ bands and oxygen $3p$ states, changing the physics qualitatively.

In the cuprates all of the d orbitals are filled except d$_{x^2-y^2}$ and we measure $N_d$ with respect to the $d^8$ core   so $0\leq N_d \leq 2$ with the physical range being $1\leq N_d\leq 2$.  In the nickelates all d-orbitals are filled except the $d_{x^2-y^2}$ and $d_{3z^2-r^2}$  and we measure $N_d$ with respect to the $d^6$ core so $0\leq N_d\leq 4$ with the physical range being $1\leq N_d\leq4$.  The  results  are seen to have a remarkably simple structure. For physically relevant $U$ values the phase boundary is essentially vertical and occurs at $N_d\sim1.2-1.25$ in the two dimensional (cuprate) case and  at $N_d\sim 1.25-1.35$ in the three dimensional  (nickelate) case.    The simplicity of the $U-N_d$ phase diagram  identifies $N_d$ as a  critical variable for placing transition metal oxides on the metal-insulator phase diagram.  Further evidence is that the VASP/Wannier (red empty square) and WEIN2K/projector (green filled circle) calculations  yield essentially the same cuprate phase diagram  although the orbitals are defined differently and the band theory $N_d$ are in fact slightly different.   These results are consistent with  calculations of Ref.~\onlinecite{Wang11} which show that the details of the oxygen bandstructure  do not affect the location of the metal-insulator transition if the  $N_d$ are tuned to be the same also support this conclusion. 

We note that the band theory values of $N_d$ are $\sim2.16$ (nickelates) and $\sim 1.45$ (cuprates). The very large difference observed in the nickelate case between the band theory $N_d$ and the value needed to drive a Mott/charge-transfer transition raises questions about the relevance of conventional Mott/charge-transfer physics to the nickelates. The band theory $N_d$ in the cuprates is closer to the phase boundary (although still outside it), suggesting a more important role for Mott physics in these materials.

\begin{figure}[t]
\begin{center}
\includegraphics[angle=0.0, width=0.9\columnwidth]{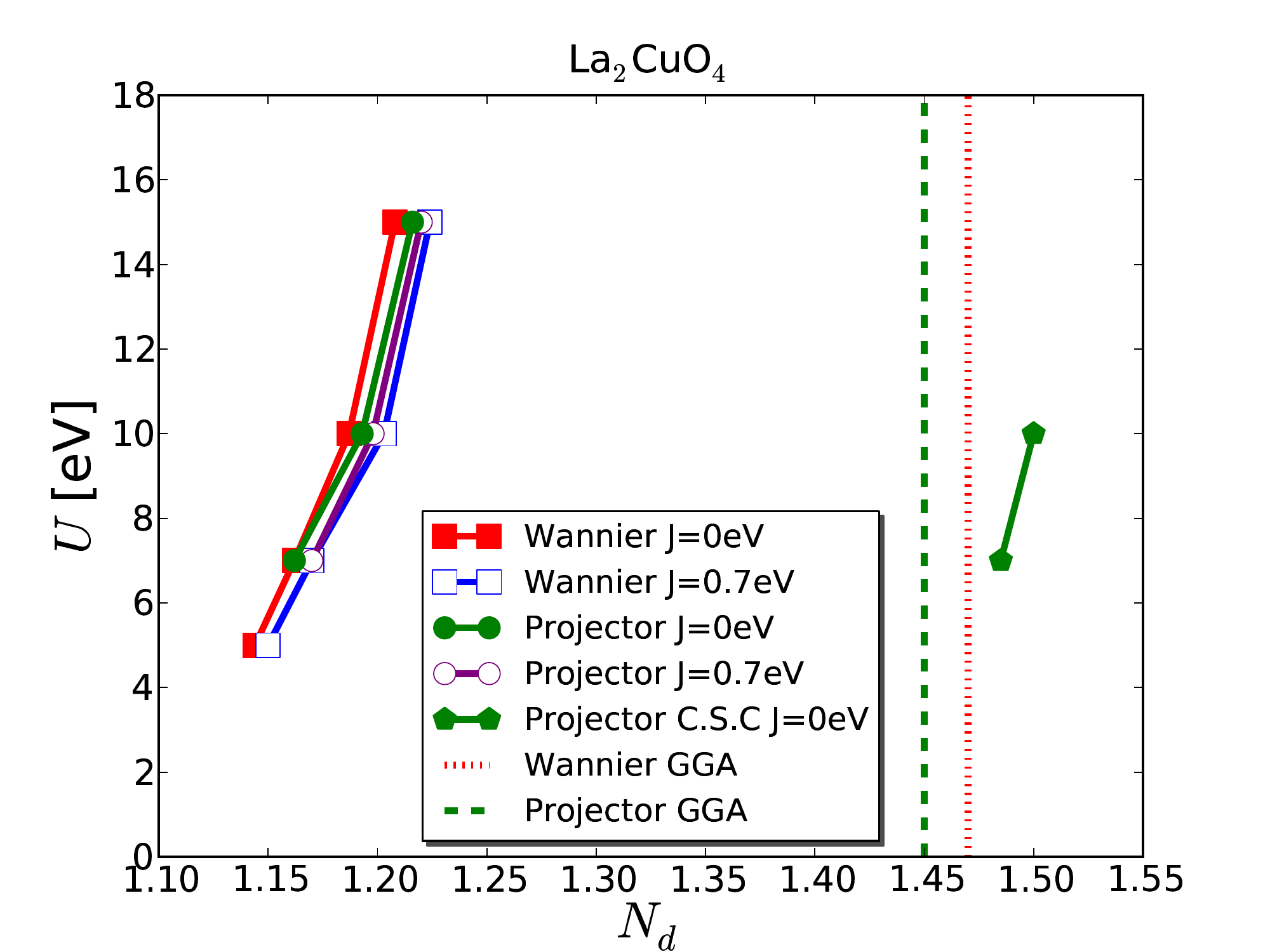}
\caption{(Color online) Metal-insulator phase diagram indicating limit of stability  of insulating phase  $N_{d,c1}$  for cuprates calculated for the one orbital model of cuprates using VASP/Wannier method (red filled square and blue empty square) and   for the five orbital model using the Wien2K/projector method  (green filled circle and purple empty circle) method at Hunds coupling values indicated.  Also indicated as vertical lines are the Wein2K/Projector and VASP/Wannier band theory estimates for the d-occupancy  and the value obtained from the fully charge self-consistent DFT+DMFT Wein2K/Projector procedure at $J$=0eV (green pentagon). }   
\label{cupratepd}
\end{center}
\end{figure}

We now turn to a more detailed examination of the metal-insulator transition phase diagram predicted for cuprates at different Hund's coupling values ($J$=0.7 and 0eV) using different calculational schemes. Fig.~\ref{cupratepd} compares the metal-insulator phase boundaries obtained using VASP/Wannier and Wein2K/projector schemes with Hunds coupling $J-0$ and $0.7eV$.  All the calculational schemes show very similar phase diagrams, with insulating behavior confined to a very narrow $N_d$ region, $N_d\lesssim1.2$. The VASP/Wannier calculation treats only the d$_{x^2-y^2}$ orbital dynamically (with DMFT), treating the other orbitals by Hartree-Fock,   while the Wien2k/projector calculation treats the full five d orbitals with DMFT (in the Ising approximation). We also verified (not shown) that use of the full 5 orbitals in the VASP/Wannier scheme does not change the phase boundary appreciably. Also presented as vertical lines in Fig.~\ref{cupratepd} are the $N_d$ values obtained from the two underlying band theory methods: these are somewhat larger than the $N-d$ required to drive an insulating phase, although in contrast to the nickelate case  the displacement of the band theory $N_d$ from the critical $N_d$ needed for insulating behavior is not large. The similarity between the results  demonstrates again that $N_d$ is the theoretically important parameter and that the results are robust to choice of computational method. 

At small $U$ one might expect that the phase boundaries of extrapolate linearly towards $U=0$, $N_d=1$ because as $N_d \rightarrow 1$ the $d$ bands become far removed from the $p$ bands so both the $d$ bandwidth and $N_d-1$ $\sim 1/\Delta$ (there will be some small d-d hopping which will make U extrapolate to some small non-zero value). We have not pushed our calculations into this regime because it is unphysical for the transition metal oxides. Accessing the small-U phase boundary by tuning parameters so that $N_d$ is moved  closer to $1$  corresponds to placing the d-bands at such a high energy relative to the p-bands  that they would be mixed with transition metal $t_{2g}$ bands, rare earth $f$ and $d$ bands and oxygen $3p$ states, changing the physics qualitatively.  

The phase diagrams  presented in Figs.~\ref{pdnd},\ref{cupratepd} are obtained by use of a fixed band structure and a varying double counting correction. We have also performed calculations for the cuprates using the fully charge self-consistent DFT+DMFT procedure implemented in Wein2K/projector method, with the $E_{dc}=U(N_d-0.5)-0.5\cdot J(N_d-0.5)$ and $U=7,10$. In this procedure the Kohn-Sham band structure is made self-consistently to the density produced by the interacting Green function. The results are shown as green pentagons in Fig.~\ref{cupratepd} and correspond to $N_d$ (1.47$\sim$1.5) slightly higher than the band theory value (1.45)  and well within the metallic regime.  The self-consistently determined Kohn-Sham band structure is only slightly changed from the DFT band structure at the same $N_d$, as will be seen from the spectral functions to be discussed below.

\subsection{Spectra}

\begin{figure}[t]
\begin{center}
    \includegraphics[angle=0.0, width=0.9\columnwidth]{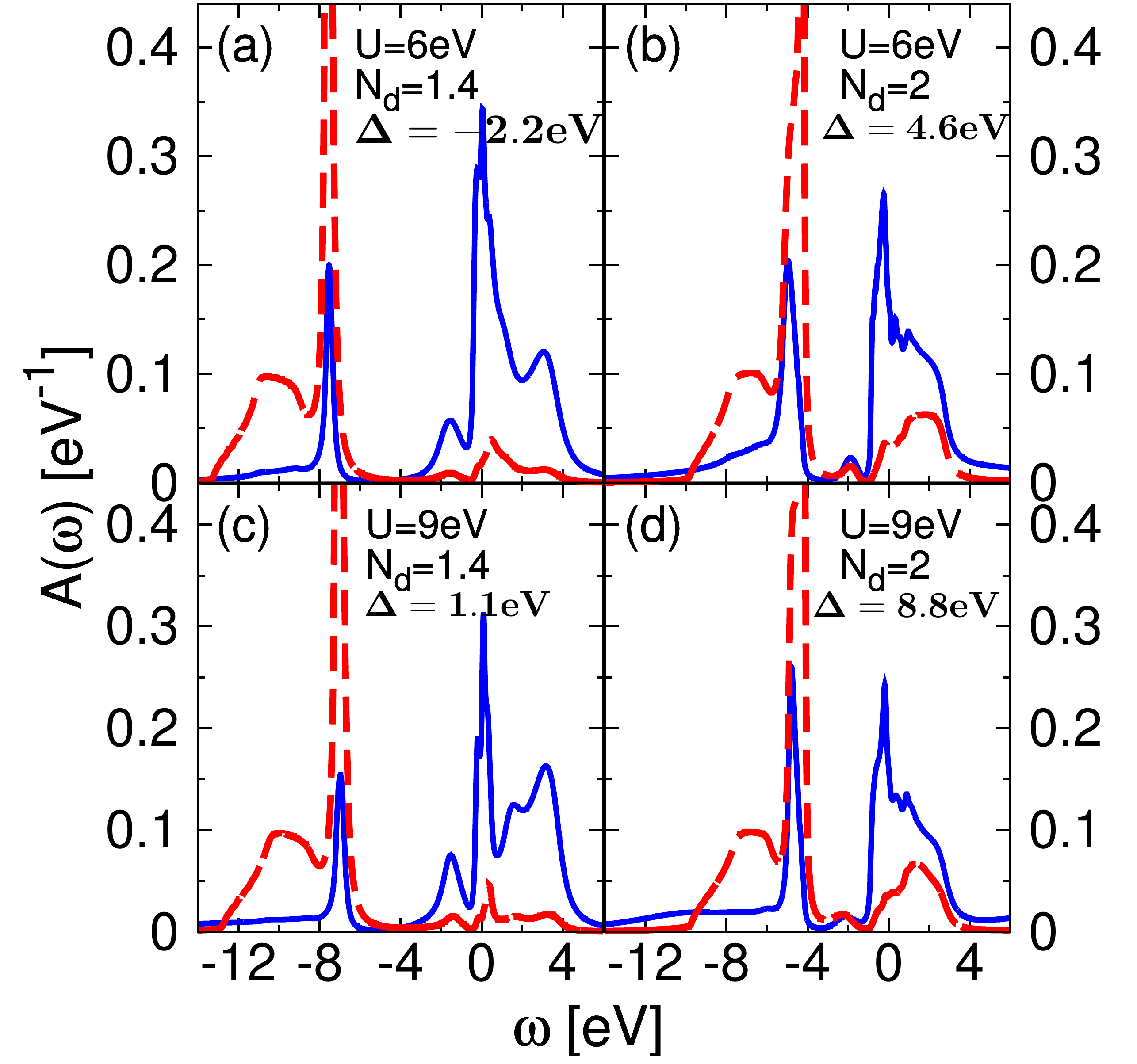}
\caption{(Color online) Local spectral functions computed  for the nickelates at different values of charging energy and bare charge transfer energy $\Delta$ (see main text for precise definition) as indicated.  Solid blue lines show the $d$-density of states and the dashed red lines show the $p$-density of states.}
    \label{DOS}
\end{center}
\end{figure}

Further insight into the role of $N_d$ may be obtained from examination of the electron spectral function (many-body density of states) obtained  by maximum entropy analytical continuation from our imaginary time QMC data using the approach in Ref. \onlinecite{Wang09}. Fig.~[\ref{DOS}] shows the spectral functions for the $ReNiO_3$ calculation  for several different combinations of interaction strength and charge transfer energy (double counting correction), here defined in terms of the energy difference $\Delta$ between  $\Gamma$-point $p_\sigma$ and $d_{x^2-y^2/3z^2-r^2}$ states.  The similarity of  spectra with similar $N_d$ but different $\Delta$ and  $U$ is evident. In particular, the distance between the fermi surface (d-dominated) peaks and the non-bonding oxygen peaks (appearing as near-delta function contributions) depend only on $N_d$. Differences of detail are evident; in particular a $U$-dependence of the {\em width} of the fermi surface quasiparticle peaks (i.e. a difference in mass)  is clearly evident in the left panels, and occurs also at the larger $N_d$ calculations shown in the right-hand panels.

\begin{figure}[t]
\begin{center}
    \includegraphics[angle=0.0, width=0.95\columnwidth]{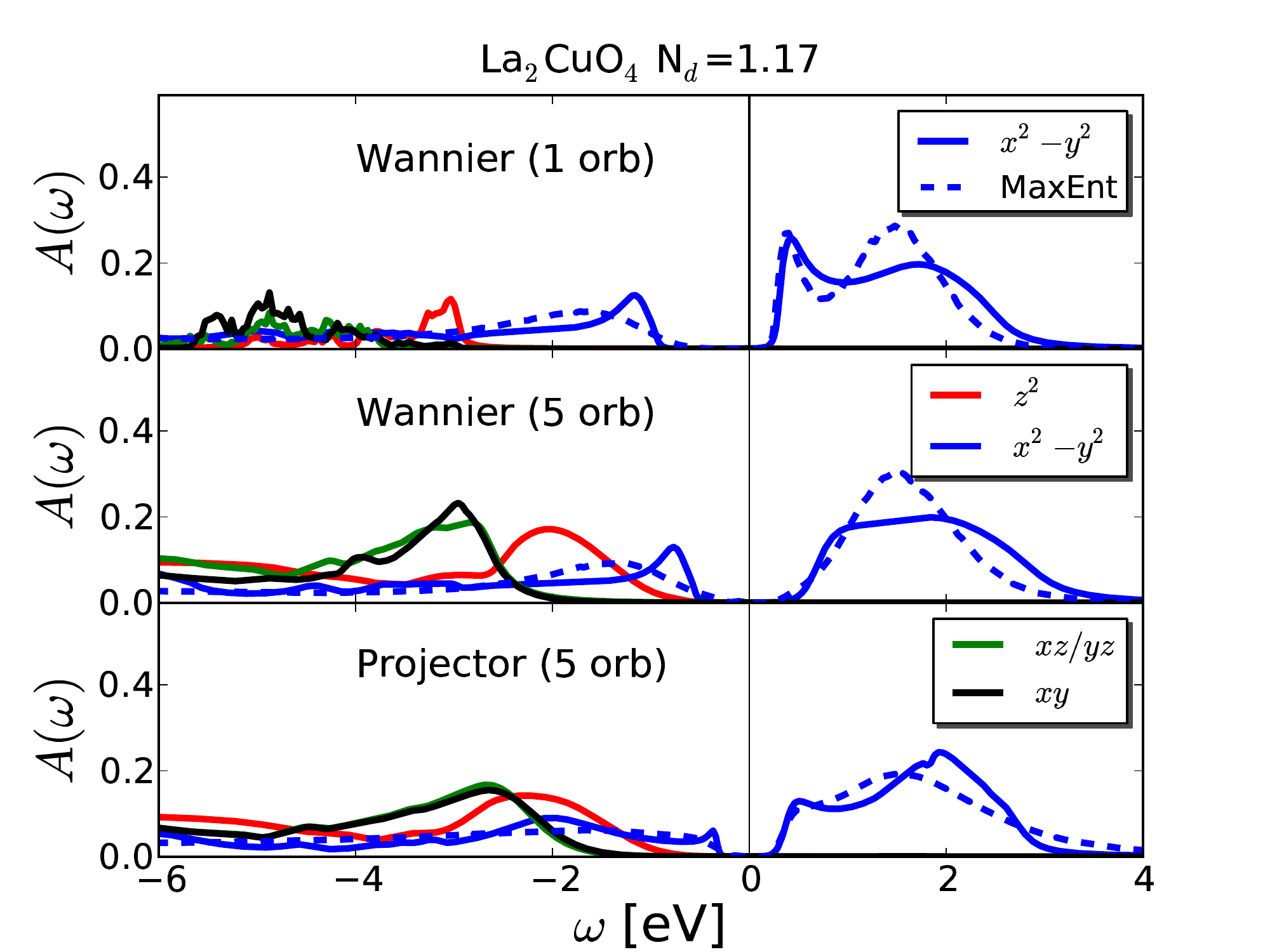}
\caption{(Color online) Local d spectral functions for the cuprates obtained for $U$=10eV and $J$=0eV from the VASP/Wannier one orbital (top) and five orbital (middle)calculations and from the Wien2k/projector five orbital  calculation (bottom). The spectral function is computed by analytically continuing self energy data using the method given in Ref.~\onlinecite{Haule:10}. The d$_{x^2-y^2}$ spectral functions obtained by maximum entropy analytical continuation  are also shown for comparison (blue dashed lines). The resulting $N_d$ values are the same ($1.17$) for all panels
.}
    \label{cuprateDOS}
\end{center}
\end{figure}

Fig.~\ref{cuprateDOS} shows the local spectral functions for $La_2CuO_4$ calculated using both the VASP/Wannier and Wien2k/projector calculations at the same $N_d$ values. The top panel shows the one orbital result of VASP/Wannier in which only d$_{x^2-y^2}$ orbital  is treated with DMFT while other d orbitals are treated by HF. Four d orbitals approximated by HF are fully filled and the spectra are concentrated mostly at low energies below -6eV.  d$_{x^2-y^2}$ orbital treated with DMFT shows a gap with the size of nearly 1eV. The middle panel shows the VASP/Wannier result for the case  in which entire five d orbitals are treated by DMFT.  The four orbitals except d$_{x^2-y^2}$  are still fully filled but the spectra are broadly distributed over the wide energy range below the Fermi energy. The spectral gap of d$_{x^2-y^2}$ orbital has a similar size as the top panel result. The bottom panel displays the five orbital result of Wien2k/projector and the filled four d orbitals show 
a  broad spectral distribution similar to the five orbital VASP/Wannier result. However, the d$_{x^2-y^2}$ orbital gap is rather smaller compared to the VASP/Wannier results. This is the only difference we have found between results computed at the same $N_d$ and is presumably due to the different definitions of correlation strength following from the different definitions of correlated orbitals. 

\begin{figure}[t]
\begin{center}
    \includegraphics[angle=0.0, width=0.9\columnwidth]{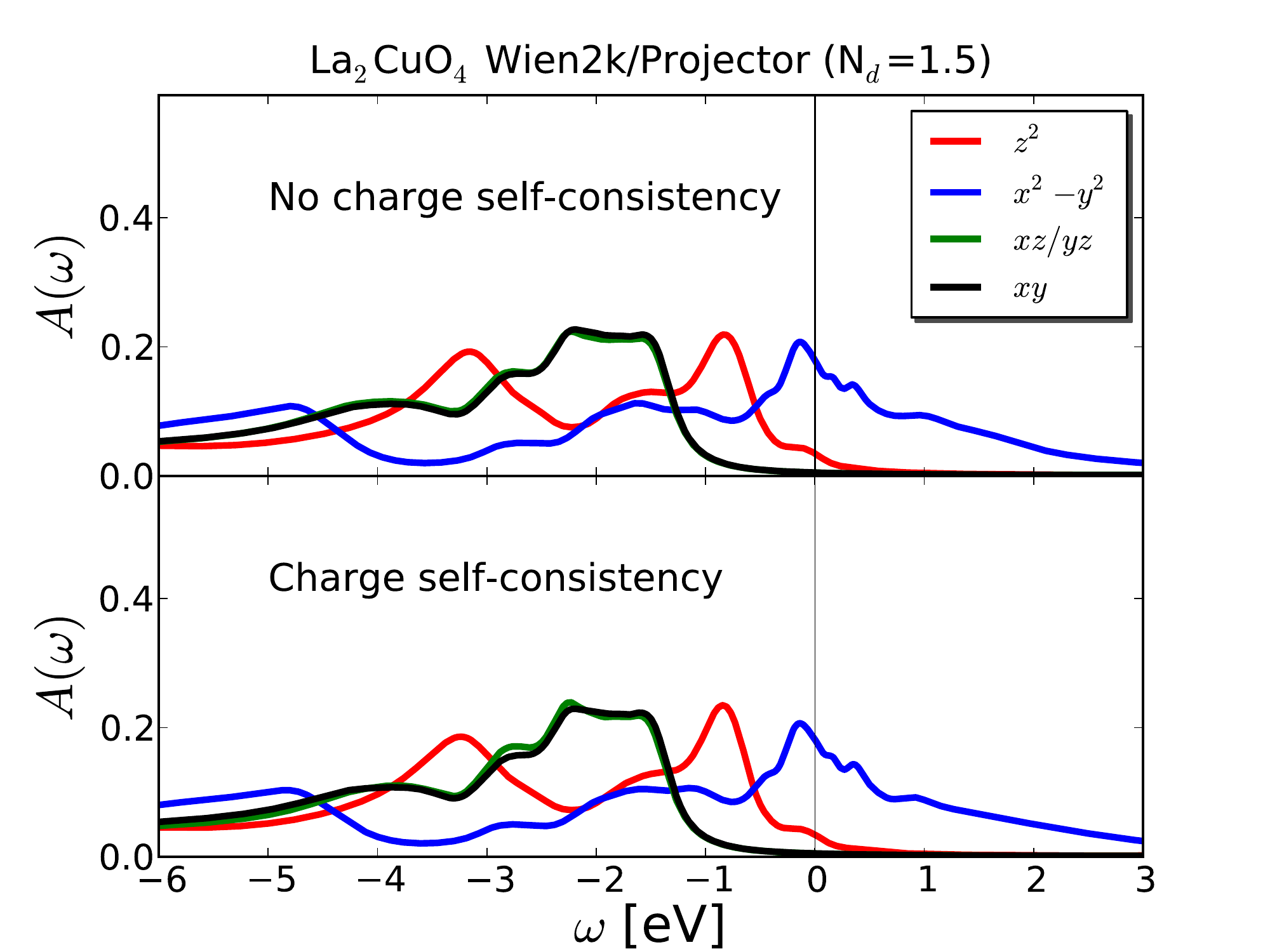}
\caption{(Color online) Local d spectral functions for the cuprates obtained  using the Wien2k/Projector method without (top panel) and with (bottom panel) full charge self-consistency. The charge self-consistent result was obtained using the standard double counting as described int he text; the double counting correction in the non-self-consistent calculation was adjusted to yield the same $N_d$. }
    \label{DOS_selfc}
\end{center}
\end{figure}

Another illustration of the key role played by $N_d$ is obtained from the comparison shown in Fig.~\ref{DOS_selfc} of the spectral functions obtained from different cuprate calculations. The lower panel shows the spectral functions obtained using the fully charge self-consistent procedure described above with the standard double counting formula, while the upper panel shows the spectral functions obtained from a DFT+DMFT calculation based on  the Wein2K/projector band theory with $E_{CT}$ adjusted so the final $N_d$ matches the fully self-consistent one. As the d$_{x^2-y^2}$ orbital gets renormalized in the self-consistent DFT+DMFT charge density, the d orbital energy levels except d$_{x^2-y^2}$ move slightly away from the Fermi energy by very small amounts of order $0.03eV$. But the basic similarity of the two calculations shows again that $N_d$ is the crucial variable and also indicates that the renormalization of the band theory by the fully charge self-consistent procedure is small.

\section{Discussion \label{Discussion}}
The generally accepted Zaanen-Sawatzky-Allen theory of late transition metal oxides describes the metal-insulator-transition physics of transition metal oxides in terms of two parameters: $U$, a charging energy associated with the transition metal d-orbitals, and $\Delta$, an energy difference between transition metal d and oxygen p levels. The quantity $\Delta$ is difficult to define precisely for two reasons. First,  the oxygen states in typical transition metal oxides are spread out into a band, so it is not clear what energy to use for $\varepsilon_p$. Second, the physical  d-level energy has a substantial contribution from the many-body interactions of interest, both directly and via the charge transfer energy $E^{dc}$.  In physical terms $\Delta$ parametrizes the degree of covalence between the ligand and transition metal orbitals. The results presented in this paper indicate that it is very useful to parametrize this covalence instead by  the $d$-occupancy $N_d$, which is a single-valued function of $\Delta$.  We show that the Zaanen-Sawatzky-Allen metal-insulator phase diagram takes a very simple, quasi-universal form when expressed in the  $U-N_d$ plane, and that the main features of the calculated many-body spectra depend mainly on $N_d$: different combinations of $U$, band parameters and $E^{dc}$ give very similar spectra if they yield the same $N_d$.  Expressing the phase diagram in the $U-N_d$ plane reveals an interesting and apparently previously unnoticed phenomenon: the Mott-Zaanen-Sawatzky-Allen insulating state is only obtained if $N_d$ is relatively close to an integer value (low covalence) and at larger $U$ the phase boundary is nearly vertical. Thus within the Zaanen-Sawatzky-Allen picture  local correlations cannot  drive a metal insulator transition unless the physical covalence is quite small.

$N_d$ is a theoretical quantity, precisely defined only within a specific calculational scheme.  The values obtained for $N_d$ depend on the method which is used, so the $U-N_d$ phase diagram is in principle specific to a given calculational method. In most of the results presented in this  paper we have defined the correlated orbitals using a maximally localized Wannier function construction \cite{Souza01,Amadon08} with an energy window taken large enough to include all of the $d-p$ complex of bands. The resulting $Ni_d$ Wannier functions are highly localized.
We have also investigated other prescriptions, for example the projector methods used in the VASP DFT+U and the projector-based DFT+DMFT methods~\cite{Haule:10}. We find that if an adequate energy window is used, all of the procedures employed in current literature give very similar answers (for examples, see  Figs.~\ref{pdnd},\ref{cupratepd}). 

Although $N_d$ is a model-dependent quantity, it is interesting to compare  the values of $N_d$ obtained by different methods and to use the results  to place materials on the metal-insulator phase diagram. Density functional band theory methods (in combination with a definition of d orbital) yield predictions for $N_d$ which are indicated in Figs.~\ref{pdnd},\ref{cupratepd}) for $La_2CuO_4$ and $LaNiO_4$. The theoretical status of these results requires some discussion. If the exact density functional were known and the Kohn-Sham equations could be solved exactly, the exact density $\rho(r)=\int d\omega ImG(r,r,\omega)/pi$ would be obtained \cite{Jones89}. However, the exact $N_d$ for the transition metal centered at position $R$ is given in terms of the defined $d$ wavefunctions $\phi_d(r)$ via Eq.~\ref{Nddef}. This cannot be derived directly from any projection of   $\rho(r)$ and as a matter of principal need not be given exactly even by exact density functional theory. However, the small size of the atomic orbitals along with the general success of DFT methods in obtaining charge densities and basic band structures \cite{Jones89} suggests that the DFT estimates for $N_d$ may be reasonable. It is important to point out that there is at present no solid-state  benchmark for the quantitative reliability of the results of present state of the art functionals and some experimental results indicate that density functional
methods overestimate covalency  \cite{Dudarev00}.   Thus it might be that density functional results for $N_d$ are substantially in error in correlated materials. A systematic evaluation of the reliability of DFT estimates of transition metal-ligand covalency is also indicated.

An important question for beyond DFT calculations is the extent to which many body effects (including the double counting correction) change $N_d$ from the band theory value. In many cases the standard fully charge self-consistent implementations of DFT+U and DFT+DMFT change $N_d$ only slightly from the band theory value. For example, our recent studies of $LuNiO_3$ \cite{Park12} reveal a GGA values of  $N_d=8.21/8.20$ for the two inequivalent $Ni$ sites, whereas GGA+U with $U=5$ and $J=1$ gives $N_d=8.24/8.22$. An example for cuprates is shown in Fig.~\ref{cupratepd}. However, other calculations use a double counting procedure which gives a larger change.  For example Ref.~\onlinecite{Weber10} modeled $La_2CuO_4$ using a double counting correction which shifted $N_d$ from $\sim 1.65$ to $\sim 1.1$.  Determination of the most correct double counting correction is an important open problem in materials theory; the results of this paper suggest that the issue may usefully be addressed by consideration of the interaction-induced shifts in $N_d$.

Experimental estimation of the $d$ valence is  also of interest. Indirect measurements, for example of the transferred hyperfine couplings \cite{Walstedt01} or the  magnetic form factor in ordered states \cite{Walters09} have indicated very strong covalency effects in cuprates. The
estimates given in these papers are $N_d\sim1.6$, somewhat larger than the band theory values using well localized orbitals.  A more direct approach would be to detect  the density of holes on oxygen  via resonant X-ray scattering techniques as in Refs.~\onlinecite{Peets09,Ament10}. As indicated in Fig. ~\ref{DOS}, spectroscopic measurements of the relative position of $p$ and $d$ features in the many-body density of states would also be revealing.  One particularly informative experiment  would be a measurement of  changes in the density across the $ReNiO_3$ series, to reveal the change in $O$-hole density.

Finally, we discuss some consequences of our results. We first observe that if one chooses a double-counting which fixes  $N_d$ at the DFT value, then both $La_2CuO_4$ and $LaNiO_3$ would, within single-site DMFT, be predicted to be metals. The nickelate estimate $N_d\sim2$  is so far from the value $\approx 1.3$ needed to drive a Mott/charge-transfer metal insulator transition that it seems likely that standard Mott/charge transfer physics is simply not relevant to the rare earth nickelate family of materials.  $LaNiO_3$ is a moderately  correlated metal in experiment \cite{Ouellette10}, but  replacing $La$ by other rare earths drives a metal-insulator transition \cite{Torrance92,Alonso99} sometimes interpreted in the Mott/charge transfer paradigm \cite{Imada98,Stewart11}. We have verified that within GGA band theory (and using the Wannier prescription defined in this paper)  the d-occupancy is nearly the same for strongly insulating $LuNiO_3$ in the experimental structure as it is for $LaNiO_3$. In fact,  the insulating members of the rare-earth nickelate series are characterized by a large amplitude lattice distortion \cite{Alonso99,Medarde09} which plays a key role in the insulating behavior \cite{Park12}.  

Band structure calculations place high-$T_c$ cuprates rather closer to the metal-insulator transition than are the rare-earth nickelates, but still on the metallic side of the phase boundary. It is therefore not unreasonable that the correct double-counting correction will move the materials across the phase boundary into the insulating state (as assumed e.g. in Ref.~\cite{Weber10}). Alternatively, physics beyond the simple Mott/charge-transfer picture may  drive the insulating state. For example, cluster dynamical mean field and other calculations in the Hubbard model  at moderate interaction strengths  reveal a crucial effect of intermediate-ranged antiferromagnetic correlations \cite{Tremblay06,Gull08,Park08,Werner09,Gull10}.

More broadly our findings  raise an important issue in current materials theory.  In a beyond-DFT calculation one chooses a subset of orbitals which are subject to additional correlations. The point of view expounded by Zaanen, Sawatzky and Allen \cite{Zaanen85}, implemented in most DFT+U and DFT+DMFT codes \cite{Held06,Kotliar06} and used here is that for transition metals the correlated orbitals are  atomic-like   transition metal d orbitals; and that the remaining states (in particular the $O$-p orbitals), are included in the calculation but with correlations treated only on the DFT level. We refer to this approach, somewhat imprecisely, as the ``p-d model'' approach.  An alternative, also widely adopted in the literature, is that one should define the correlated orbitals by  downfolding the band structure to include only the frontier orbitals (in the transition metal oxide context, the p-d antibonding bands which cross the fermi level), which are then treated as  a multiorbital Hubbard model. The two approaches can lead to different physics. For example, if the total number of electrons per unit cell is an odd integer, increasing U in the multiorbital Hubbard model always leads to a metal-insulator transition, whereas in the p-d model increasing $U$ may or may not lead to a  metal-insulator transition, depending on the covalency.  Different results are also found for orbital polarization in Ni-based oxide superlattices \cite{Hansmann09,Han11}. Which approach provides the most reasonable modelling of actual materials is not completely clear, although the fact that as presently implemented the DFT+DMFT approximation captures only local correlations suggests that one should favor approaches based on well localized orbitals \cite{Kotliar06}. Further investigations are warranted into  the questions of which orbitals to correlate and of how to downfold the fully interacting model to an effective low energy model which can be studied numerically.

{\it Acknowledgements:}  AJM was supported by the US Department of Energy under grant DOE-ER-046169, MJH and CAM by the U. S. Army Research Office  via grant No. W911NF0910345 56032PH, X.W. in part by the Nanoscale Science and Engineering Initiative of the National Science Foundation under Award Number CHE-0641523 and by the New York State Office of Science, Technology, and Academic Research (NYSTAR) in part by NSF-DMR-1006282 and in part by the Condensed Matter Theory Center of the University of Maryland; LdM was supported by Program ANR-09-RPDOC-019-01 and by RTRA Triangle de la Physique. Part of this research was conducted at the Center for Nanophase Materials Sciences, which is sponsored at Oak Ridge National Laboratory by the Division of Scientific User Facilities, U.S. Department of Energy. The CT-HYB impurity solver is based on a code  developed by P. Werner, using the ALPS library \cite{ALPS}.


\end{document}